\documentclass[a4paper]{aa}

\usepackage{graphicx}
\usepackage{txfonts}
\usepackage{natbib}
\bibpunct{(}{)}{;}{a}{}{,}  

\begin{document}

\title{Seismology of Procyon A: determination of mode frequencies, amplitudes, lifetimes, and granulation noise}

\author{S. Leccia \inst{1} \and H. Kjeldsen \inst{2} \and A. Bonanno \inst{3} \and R. U. 
Claudi \inst{4} \and R. Ventura \inst{3} \and L. Patern\`o \inst{5}}

\offprints{S. Leccia; \email leccia@na.astro.it}

\institute{I.N.A.F. -- Astronomical Observatory of Capodimonte, Salita Moiariello 16, I-80131 Napoli, Italy 
\and
Department of Physics and Astronomy, University of Aarhus, Ny Munkegade, Building 1520, DK-8000 Aarhus C, Denmark 
\and
I.N.A.F. -- Astrophysical Observatory of Catania, Via S. Sofia 78, I-95123 Catania, Italy 
\and
I.N.A.F. -- Astronomical Observatory of Padua, Vicolo Osservatorio 5, I-35122 Padova, Italy 
\and     
Department of Physics and Astronomy, Astrophysics Section, University of Catania, Via S. Sofia 78, I-95123 Catania, Italy}

\titlerunning{Seismology of Procyon A}
\authorrunning{S. Leccia, H. Kjeldsen, A. Bonanno et al.}

\date{Received / Accepted }

\abstract{The F5 IV-V star Procyon A ($\alpha\!-\!\rm CMi$) was observed in January 2001 by means of the high resolution  spectrograph SARG operating with the TNG $3.5\,\rm m$ Italian telescope (Telescopio Nazionale Galileo) at Canary Islands, exploiting the iodine cell technique. The time-series of about 950 spectra carried out during 6 observation nights and a preliminary data analysis were presented in \citet{Claudi}. These measurements showed a significant excess of power between $0.5$ and $1.5\,\rm mHz$, with $\simeq 1\,\rm m\,s^{-1}$ peak amplitude. Here we present a more detailed analysis of the time-series, based on both radial velocity and line equivalent width analyses. From the power spectrum we found a typical $p$-mode frequency comb-like structure, identified with a good margin of certainty $11$ frequencies in the interval $500-1400\,\rm\mu Hz$ of modes with $l=0,1,2$ and $7\leq n \leq 22$, and determined large and small frequency separations, $\Delta\nu = 55.90\pm 0.08\,\mu\rm{Hz}$ and $\delta\nu_{02}=7.1\pm 1.3\,\rm \mu Hz$, respectively. The mean amplitude per mode ($l=0,1$) at peak power results to be $0.45\pm 0.07\,\rm m\,s^{-1}$, twice larger than the solar one, and the mode lifetime $2\pm 0.4\,\rm d$, that indicates a non-coherent, stochastic source of mode excitation. Line equivalent width measurements do not show a significant excess of power in the examined spectral region but allowed us to infer an upper limit to the granulation noise.
 
}
   
\maketitle

\section{Introduction \label{INTRODUCTION}}
Procyon A ($\alpha\!-\!\rm CMi$, HR 2943, HD61421), hereinafter simply called Procyon, is a F5 IV-V star with $\rm V=0.363$ at a
distance of $3.53\,\rm pc$ in a $40\,\rm y$ period visual binary system, the companion being
a white dwarf more than $10\,\rm mag$ fainter. By adopting the
very accurate parallax measured by HIPPARCOS, $\Pi = 285.93\pm 0.88\,\rm mas$,
\citet{prieto02} derived a mass of $1.42 \pm 0.06\,\rm M_{\odot}$, a radius
of $2.071 \pm 0.02\,\rm R_\odot$, and a gravity $\log g = 3.96 \pm 0.02$.

Procyon, owing to its proximity and brightness, has already
attracted attention of stellar seismologists \citep{brown91, guenther93, barban99,
chaboyer99, martic99, martic04, egge04}, causing an intense debate among the scientists. The excess of power in the range $0.5 - 1.5\,\rm mHz$ found by \citet{bouchy04}, \citet{martic04},
\citet{egge04}, and \citet{Claudi}, hereinafter referred as Paper I, seems to have a stellar origin and is consistent with
a $p$-mode comb-like pattern. 

However, data from the Canadian
MOST satellite \citep{matthews04} show no significant power excess in the same spectral region. To this regard \citet{bedding05} suggested that the most likely explanation
for the null detection could  be a dominating non-stellar noise source in the MOST data, though \citet{regulo05} claimed for the presence of a signal in these data.

In Paper I we presented high precision radial velocity measurements 
carried out during 6 observation nights 
by means of the high resolution spectrograph SARG operating with TNG $3.5\,\rm m$ Italian telescope (Telescopio Nazionale Galileo) at Canary Islands. The data showed an excess of power
between $0.5-1.5\,\rm mHz$ with a large separation of about $56\pm 2\,\mu\rm Hz$, in agreement with previous measurements of \citet{mosser98}, \citet{barban99}, \citet{martic99, martic04} and \citet{egge04}, who found values in the range $53 - 56\,\rm \mu Hz$.
Here we deal with an improvement of the preliminary analysis presented in Paper I concerning an accurate time-series analysis of both radial velocities and line equivalent width measurements, from which mode frequencies, amplitudes and lifetimes are determined and constraints on stellar granulation noise are deduced.

\section{Observations \label{OBSERVATIONS}}
SARG is a high resolution cross dispersed echelle
spectrograph \citep{gratton01} which operates in both single object and long slit (up to $26^{\prime\prime}$)
observing modes and covers a spectral wavelength range from $370\,\rm nm$ up to about $1000\,\rm nm$, with a resolution ranging from $R=29,000$ up to $R=164,000$.
Our spectra were obtained at $R=144,000$ in the wavelength range between 
$462 - 792\,\rm nm$. The calibration iodine cell works only in the blue part of the spectrum
($462 - 620\,\rm nm$) that has been used for measuring Doppler shifts. 
The red part of the echelle spectrum of Procyon has been used for measuring line equivalent widths of absorption lines sensitive to temperature. 
During the observing period we collected about 950 high signal-to-noise ratio (S/N) spectra with a mean exposure time of about $10\,\rm s$ (see Paper I for more details).
The red part spectrum has been calibrated in wavelength by using a Th-Ar lamp.
The analysis of both the blue and red parts of the spectrum was performed by using the IRAF package facilities.
  
\section{Radial velocity measurements and analysis \label{RADIAL}}
\begin{figure}
\centering
\includegraphics[width=8cm]{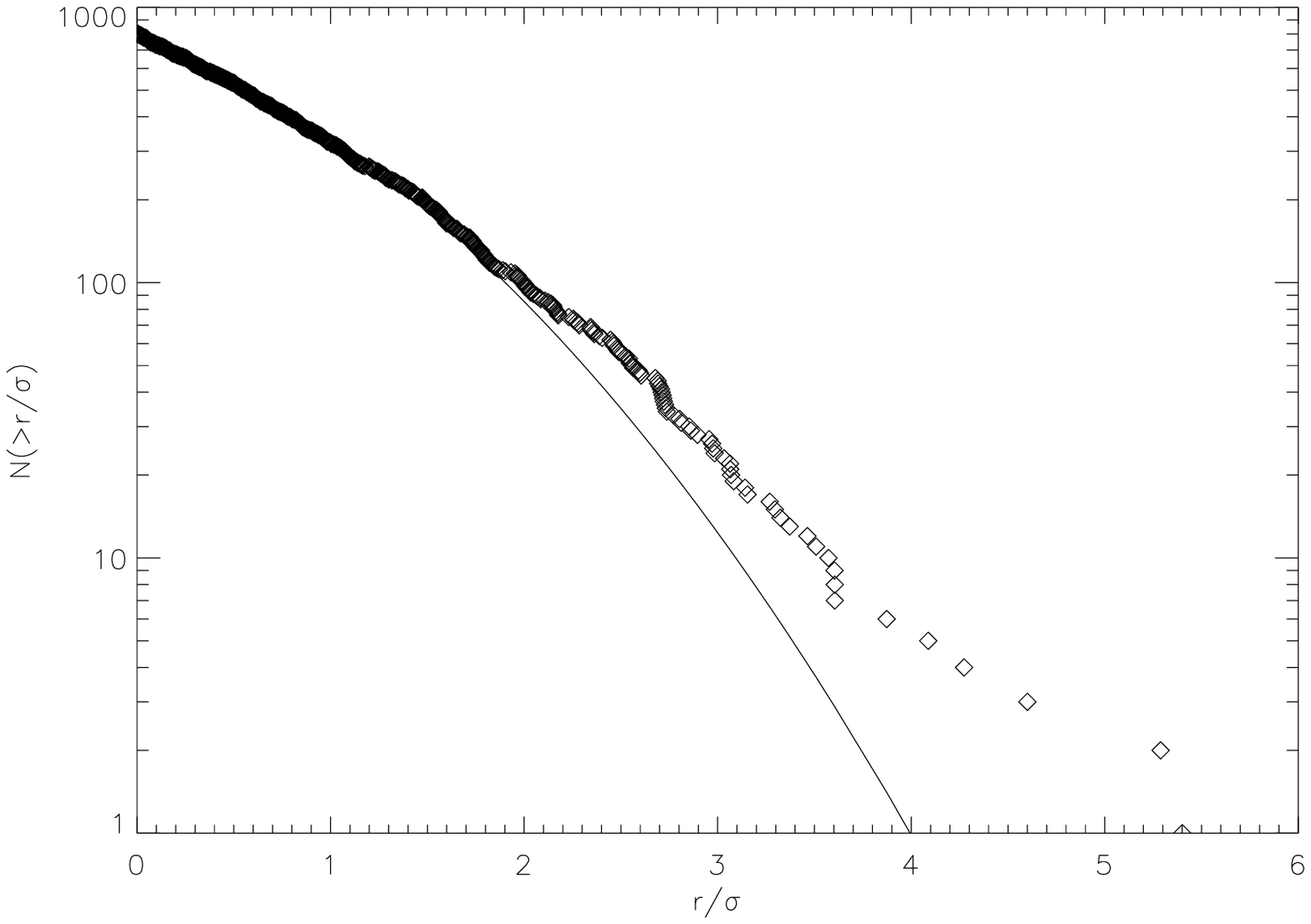}
\includegraphics[width=8cm]{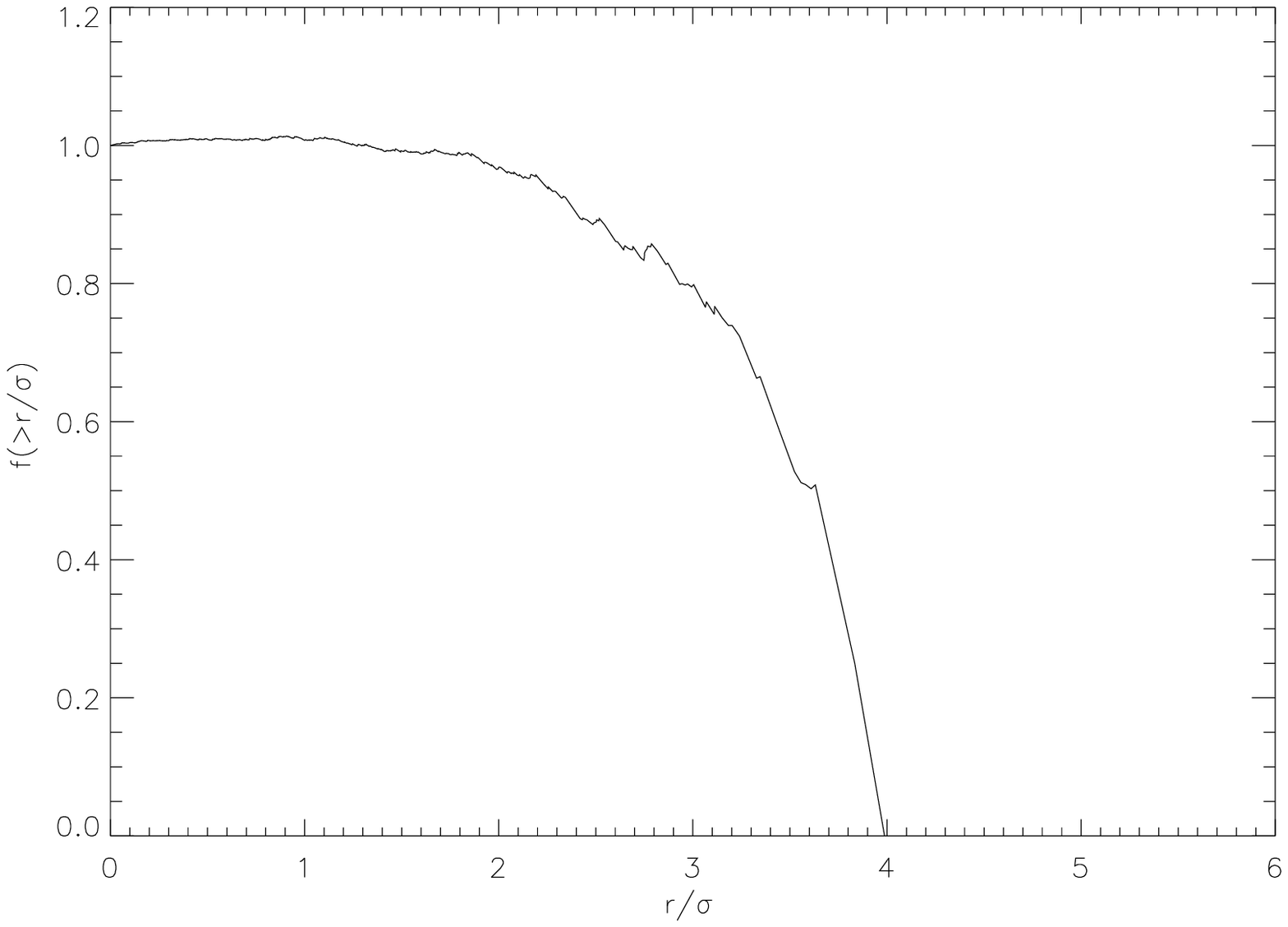}
\caption{\emph{Upper panel}: cumulative histograms of $|r_i/\sigma_i|$ for SARG data. The diamonds show the observed data, and the solid curve shows the result expected for a Gaussian-distributed noise. \emph{Lower panel}: ratio of the observed to the expected histograms, indicating the fraction of ``good'' data points.}
\label{isto}
\end{figure}

\begin{figure}
\centering
\includegraphics[width=8cm]{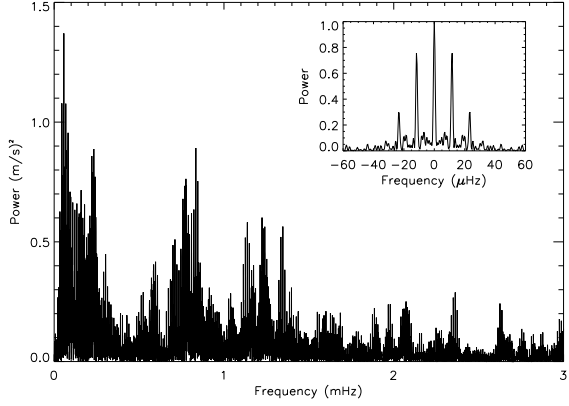}
\caption{ The power spectrum of the weighted data. An excess of power around $1\,\rm mHz$ is evident. The inset shows the power spectrum of the window function for a sine-wave signal of amplitude $1\,\rm m\,s^{-1}$, sampled in the same way as the observations. The power units are the same as those of the main figure.}
\label{power}
\end{figure}

Radial velocities have been determined by means of the AUSTRAL code \citep{endl00} which
models instrumental profile, stellar and iodine cell spectra in order
to measure Doppler shifts (see also Paper I). This code also provides an estimate of the uncertainty in the
velocity measurements, $\sigma_i$. These values have been derived from the scatter of velocities measured from many ($\simeq 100$), small ($\simeq 2\,$\AA) segments of the echelle spectrum.
In order to perform a weighted Fourier analysis of the data, we firstly verified that these $\sigma_i$ values reflected the noise 
properties of the velocity measurements, following  \citet{butler04} approach.
The high frequency noise in the power spectrum (PS) well beyond the stellar
signal, reflects the properties of the noise in the data, and
since we do expect that the oscillation signal is the dominant cause of variations in the velocity time series, we need to remove it before to analyze the noise. We do that iteratively by finding the strongest peak in the PS of the velocity time-series and subtracting the corresponding sinusoid from the time-series. This procedure is carried out
for the strongest peaks in the oscillation spectrum in the frequency range $0 - 2\,\rm mHz$, until the spectral leakage into high frequencies from the remaning power is
negligible. Thus we have a time series of residual velocities, $r_i$, that
reflects the noise properties of the measurements. We then analyze the ratio $r_i/\sigma_i$, which is expected to be Gaussian-distributed, so that the outliers correspond to the suspected data points. The cumulative histogram of $|r_i/\sigma_i|$ is shown in the upper panel of 
Fig. \ref{isto}. 
The solid curve shows the cumulative histogram for
the best-fit Gaussian distribution. A significant excess of
outliers is evident for $|r_i/\sigma_i|\ge 2$. The lower panel of Fig. \ref{isto} shows the ratio
of the values of the observed points to the corresponding ones of the Gaussian curve, i.e. the fraction $f$ of data points that
could be considered as ``good'' observations, namely those which are close to the unity. The quantities $w_i=1/(\sigma_i^2 f)$ have been adopted as weights in the computation of the weighted
PS shown in Fig. \ref{power}, where the most prominent peak has an amplitude of $\simeq 1\,\rm m\,s^{-1}$. 

\subsection{Search for a comb-like pattern \label{COMBPATTERN}}
\begin{figure}
\centering
\includegraphics[width=8cm]{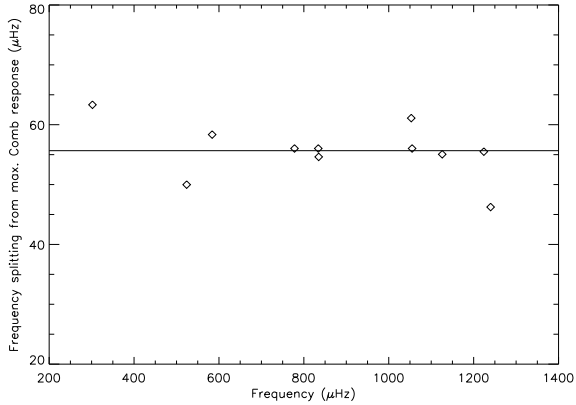}
\caption{The first-order spacing calculated over several frequency ranges, as determined from the comb-response analysis. The line shows the mean
value of $55.7\,\rm\mu Hz$, with a standard error of $\pm 1.4\,\rm\mu Hz$.}
\label{Comb}
\end{figure}

\begin{figure}
\centering 
\includegraphics[width=8cm]{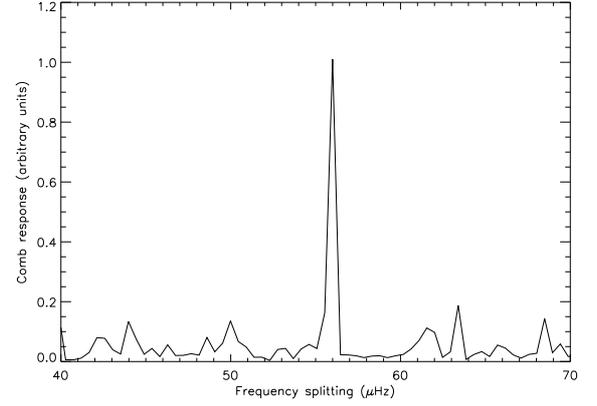}
\caption{The cumulative comb-response obtained as the sum of the individual comb-responses for each central frequency $\nu_{0}$. The peak is centred at $\Delta\nu=56 \pm 1\,\rm\mu Hz$. The error is the FWHM of the peak.}
\label{CombResponse}
\end{figure}

In solar-like stars, $p$-mode oscillations of low harmonic degree, $l$, and high radial order, $n$, are expected to produce in the PS a
characteristic comb-like structure with mode frequencies
$\nu_{n,l}$ reasonably well approximated by the Tassoul's simplified asymptotic relationship \citep{tassoul}:
\begin{equation}
\nu_{n,l}\simeq\Delta\nu(n+{l}/{2}+\varepsilon)-l(l+1)\delta\nu_{02}/6      
\label{AsymptoticRel}
 \end{equation}
where  $\Delta\nu=<\nu_{n,l}-\nu_{n-1 ,l}>$ and $\delta\nu_{02}=<\nu_{n,0}-\nu_{n-1,2}>$ are the average large frequency separation and small frequency separation for $l=0,2$, respectively, and 
$\varepsilon$ is a constant of the order of unity, sensitive to the sound speed near the surface layers of the star.
$\Delta\nu$ reflects the average stellar density, while  $\delta\nu_{02}$ is largely determined by the conditions in the core of the star and reflects its evolutionary state. 
In order to estimate the large frequency separation $\Delta\nu$ in the region of the excess of power detected in the PS, we applied 
the comb-response method, that is a generalization of the PS of a PS and consequently allows us to search for any regularity in a spectral pattern. The comb-response function, CR($\Delta\nu$), is defined by \citet{kjeldsen95} as the following products of PS calculated for a series of trial values $\Delta\nu$:
\begin{eqnarray}
\rm CR(\Delta\nu)=&\rm PS(\xi_{-1/2})PS(\xi_{+1/2})PS(\xi_{-1})PS(\xi_{+1})\times \nonumber \\
&\left[\rm PS(\xi_{-3/2})PS(\xi_{+3/2})PS(\xi_{-2})PS(\xi_{+2})\right]^{1/2}
\label{CombResponseFormula}
\end{eqnarray}
where $\xi_{\pm k}=\nu_{\rm max}\pm k\Delta\nu$, with $\nu_{\rm max}$ the  frequency of the strongest peak in the PS.
A peak in the CR at a particular spacing $\Delta\nu$ indicates the presence of a regular series of peaks in the PS, centred at the central frequency $\nu_{\rm max}$ -- tentatively assumed to be an oscillation mode -- and having a spacing $\Delta\nu/2$.
In order to select the central frequencies $\nu$'s  for the CR analysis we evaluated the white noise in the PS between $2.5-3.0\,\rm mHz$ and
adopted as $\nu_{\rm max}$'s the frequencies of the peaks in the PS between $0.5-1.5\,\rm mHz$ with $\rm S/N > 4$.
For each central frequency $\nu_{\rm max}$  we searched for the maximum CR in the range $20\leq 
\Delta\nu \leq 80\,\rm \mu Hz$. Figure \ref{Comb} reports the $\Delta\nu$'s determined from the CR analysis, that gives a mean value $\Delta\nu = 55.7 \pm 1.4\,\rm\mu Hz$, where the error is the standard deviation, and Fig. \ref{CombResponse} the cumulative CR function computed as the sum of all the individual CR functions referring to each central frequency $\nu_{\rm max}$, that gives $\Delta\nu = 56\pm 1\,\rm\mu Hz$, where the error is the FWHM of the peak.
 
\subsection{Oscillation frequencies and mode amplitude \label{OSCILLATION}}
\begin{figure}
\centering
\includegraphics[width=8cm]{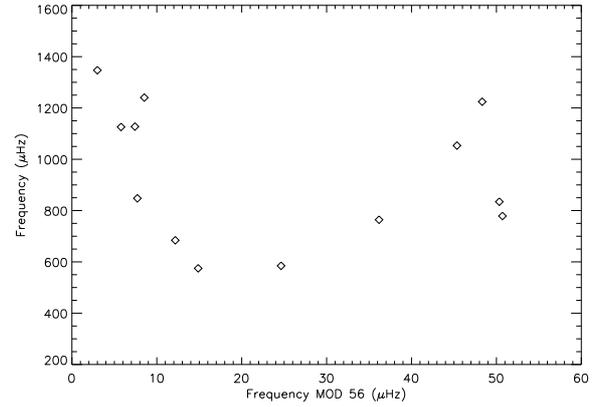}
\caption{The echelle diagram as derived from the ``standard'' extraction of frequencies.}
\label{Echelle}
\end{figure}

\begin{figure}
\centering
\includegraphics[width=8cm]{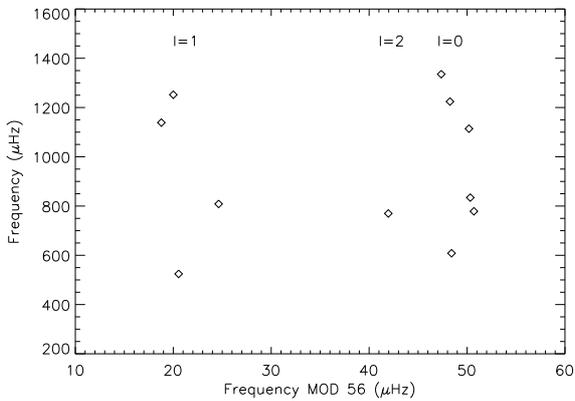}
\caption{The echelle diagram as derived from the ``modified'' extraction of frequencies. }
\label{Echelle1}
\end{figure}

\begin{figure}
\centering
\includegraphics[width=8cm]{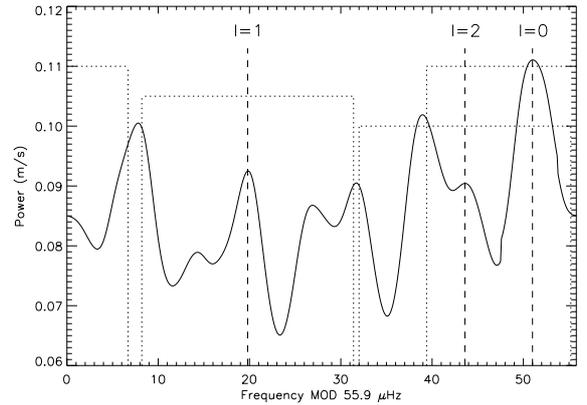}
\caption{The mean power folded at the large separation frequency as a function of frequency modulo of $55.90\,\rm\mu Hz$. The positions of the modes with $l=0,1,2$ are indicated together with the positions of the daily side-bands.}
\label{foldedspectrum}
\end{figure}

By assuming the $56\,\rm\mu Hz$ regularity as genuine, we attempted to identify the oscillation frequencies directly from the PS. In order to extract the frequencies we used the following procedure (``standard'' extraction):
i) we found the largest peak, in the frequency range $0.5-1.5\,\rm mHz$, then subtracted the corresponding sine-wave function from the
time-series, finally recomputed the PS; ii) we repeated such a procedure
for the new largest peak in the same range of frequency,
the procedure being stopped when there were no peaks with amplitude larger
than $0.29\,\rm m^2\,s^{-2}$, namely larger than $3\sigma$,
where $\sigma$ is the white noise evaluated in the PS
between $2.5-3\,\rm mHz$.
The extracted frequencies are plotted in the echelle diagram shown in
Fig. \ref{Echelle}. 
Owing to single site observations, we need to take into account the daily alias of $11.57\,\rm\mu Hz$ (see the spectral window in Fig. \ref{power}) in order to identify the modes in the echelle diagram; in fact, we cannot know
{\it a priori} whether or not the frequencies, selected by the iterative procedure described above, are the correct ones or just aliases.
Therefore, instead of shifting the frequencies by the daily alias,
that, for each shift, might lead to an arbitrary identification of the
modes, we prefer to use a procedure (``modified'' extraction) which relies on two assumptions only: i) the largest peak in the range of interest is a true pulsation mode; ii) the large separation value is $56\,\rm\mu Hz$.
We operated in the same way as in the point i) of the ``standard'' extraction, then defined a new searching spectral
region centred at a distance of $56\,\rm\mu Hz$ from the first component, the extension of which was $2\,\rm\mu Hz$, a suitable frequency span related to spectral resolution.
If this region contained a peak with an amplitude larger than $3\sigma$,
we identified this peak as the second component to be cleaned, and
recomputed the PS.
This process was repeated, shifting frequency back and forward by multiples of $56\,\rm\mu Hz$, and once the whole region of interest had been covered, the remaining largest peak was identified and the
process restarted with a new set of searching regions separated again by
$56\,\rm\mu Hz$, the procedure being stopped when all remaining peaks were below the $3\sigma$ threshold.
This is the same procedure as that used by \citet{kjeldsen95} for $\eta\,\rm Boo$, that allows to determine the mode frequencies
except for the daily alias that could cause a shift of $\pm 11.57\,\rm\mu Hz$. The frequencies extracted with this procedure are shown in  Table \ref{FI} and the related echelle diagram in Fig. \ref{Echelle1}.
\begin{table}
\caption[]{
Prominent peaks in the power spectrum of Procyon identified
as mode oscillation frequencies (in $\rm\mu Hz$) in terms of $n$ and $l$.  The values of $n$ are deduced from the fit to the asymptotic relationship (\ref{AsymptoticRel}). The frequency indicated in parenthesis was excluded from the fit because of its large deviation from the asymtotic relationship, though the related peak in the power spectrum is a high S/N one.}
\begin{center}
\begin{tabular}{cccc}
\hline
$n$&$l=0$&$l=1$&$l=2$\\
\hline
7& &524.5 &\\
9&608.4&&\\
11&&&770.0\\
12&778.7&(808.6)&\\
13&834.3&&\\
18&1114.2&1138.8&\\
20&1224.2&1252.0&\\
22&1335.4&&\\
\hline
\end{tabular}
\end{center}
\label{FI}
\end{table}
The large and small frequency separations, and the constant $\varepsilon$ have been determined by means of a least square best fit of frequencies with the asymptotic relationship (\ref{AsymptoticRel}), obtaining: $\Delta\nu=55.90\pm 0.08\,\rm\mu Hz$, $\delta\nu_{02}=7.1\pm 1.3\,\rm\mu Hz$, $\varepsilon=1.913\pm 0.025$, with a rms scatter per mode of $1.23\,\rm\mu Hz$.
In Table \ref{FI} we also give a
possible mode with $l=1$ and $n=12$ at $808.6\,\rm\mu Hz$, however this mode 
was excluded from the calculation of the fit because of its large deviation
from the asymptotic relation, though the related peak in the power spectrum is a high S/N one.  
Once determined $\Delta\nu$, it is possible to construct a folded spectrum where the power in the PS is folded at the large separation, by using the frequency modulo of $55.90\,\rm\mu Hz$. The result is shown in Fig. \ref{foldedspectrum} where the mean power as a function of frequency modulo is plotted, and the positions of modes with $l=0,1,2$ are given together with the positions of the daily side-bands, from which the detection of a $p$-mode structure in Procyon is evident.

\begin{figure}
\centering
\includegraphics[width=8cm]{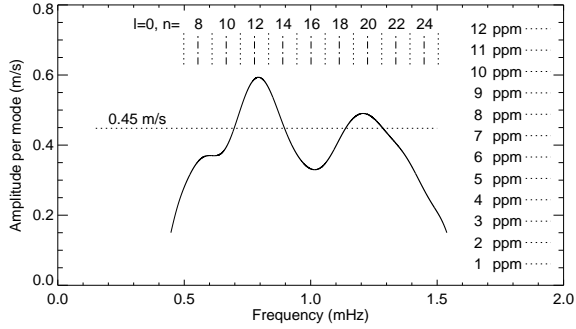}
\caption{Velocity amplitude per mode ($l=0,1$) as a function of frequency. The horizontal line at $0.45\,\rm m\,s^{-1}$ represents the mean amplitude of the $p$-mode peak power in the frequency interval $0.6-1.4\,\rm mHz$. The scale of intensity is also given together with the positions of $l=0$ modes, as derived from asymptotic relationship (\ref{AsymptoticRel}).}
\label{modeamplitude}
\end{figure}

The amplitude of individual modes can be estimated from the power concentrated in the PS, on assuming that, owing to the full disc observations, only $l=0,1,2,3$ modes are detected. It is then possible to estimate the amplitude per mode necessary for producing the observed power level. Following the procedure described in \citet{kjeldsen05} we found a mean amplitude of the $p$-mode peak power, for the modes with $l=0,1$ in the frequency interval $0.6-1.4\,\rm mHz$, of $0.45\pm 0.07\,\rm m\,s^{-1}$, that is a value twice larger than the solar one and in very good agreement with the amplitudes estimated by \citet{brown91}. Our velocity amplitude can be transformed in intensity amplitude by using the \citet{KjeldsenBedding95} method, and obtaining, at $5500\,\AA$, the value of $7.3\pm 1.1\,\rm ppm$ per mode ($l=0,1$) in good agreement with the WIRE \citep{bruntt05} and upper limit MOST \citep{matthews04}  measurements. The results are summarized in Fig. \ref{modeamplitude} where the velocity amplitude per mode is shown as a function of frequency together with the intensity scale and the equispaced positions of $l=0$ modes as derived from the relationship (\ref{AsymptoticRel}) in which the determined values of $\Delta\nu$ and $\varepsilon$ were inserted.
In Fig. \ref{modeamplitude} a dip at $1\,\rm mHz$ is clearly visible, that is a confirmation of a previously reported result of \citet{BeddingKjeldsen06}, and whose presence is probably related to the damping rate of the modes \citep{houdek99}.

\subsection{Comparison with other frequency determinations \label{COMPARISON_FRE}}
Our results are directly comparable with those of \citet{egge04}, as given in 
their Table 3, where the 
identified frequencies are tabulated. 
We may also make a comparison with the raw extracted frequencies given by
\citet{egge04} in their Table 2.
We are not able to make a direct comparison with the results by
\citet{martic04}, since they only provide a list of identified mode 
frequencies that depend on the fitting to theoretical models.

In order to perform the comparison with \citet{egge04}
we apply the same least square best fit 
we applied to our frequency data to those of \citet{egge04} and derive the 
parameters of the asymptotic relationship which result to 
be: $\delta\nu_{02}=4.9\,\rm\mu Hz$, $\varepsilon=1.537$. 
In this procedure the value of $\Delta\nu$ has been taken fixed to the 
value given by \citet{egge04}, $\Delta\nu=55.5\,\rm\mu Hz$.
If one calculate the frequencies corresponding to the two 
asymptotic solutions, the one in the present paper and the
one we obtain by fitting to the \citet{egge04} data, we find that most of the
calculated frequencies, for $l = 0, 1$, agree within $3\,\rm\mu Hz$. 
We find that
$l=0$, $n=9-24$ modes from the \citet{egge04} asymptotic relation
correspond to $l=1$, $n=8-23$ modes from the asymptotic relation in
the present study. 
We also find that $l=1$, $n=7-22$ (\citet{egge04}) modes correspond to 
$l=0$, $n=7-22$ modes identified by us. 
We conclude that we and \citet{egge04} are detecting signal
from the same underlying p-modes. 
If we look at all the raw extracted frequencies given by
\citet{egge04} in their Table 2, it appears of striking importance that most of those
frequencies may be identified by using the asymptotic solution
deduced from the present study. In table \ref{FCOMP} we compare
the \citet{egge04} raw frequencies with the asymptotic relation from
the present study ($\Delta\nu=55.90\,\rm\mu Hz$, $\delta\nu_{02}=7.1\,\rm\mu Hz$, $\varepsilon=1.913$).
\begin{table}[h]
\caption[]{
Comparison between the \citet{egge04} extracted raw frequencies 
(in $\rm\mu Hz$) and the asymptotic relation from the present
study ($\Delta\nu=55.90\,\rm\mu Hz$,
$\delta\nu_{02}=7.1\,\rm\mu Hz$, $\varepsilon=1.913$).}

\begin{center}
\begin{tabular}{cr|rccl}
\hline
Raw frq. & Correction & Fit & & & This study \\
\hline
$\nu$& &$\nu$&$l$&$n$\\
\hline
630.8 &                 &        &   &    & Noise? \\
651.5 & + 11.6 = 663.1  &  665.9 & 0 & 10 &        \\
662.7 &                 &  665.9 & 0 & 10 &        \\
683.5 & + 11.6 = 695.1  &  691.5 & 1 & 10 & ?      \\
720.6 &                 &  721.8 & 0 & 11 &        \\
791.8 & - 11.6 = 780.2  &  777.7 & 0 & 12 & 778.7  \\
797.9 &       (809.5?)  &  803.3 & 1 & 12 & 808.6? \\
799.7 &       (811.3?)  &  803.3 & 1 & 12 & 808.6? \\
828.5 &                 &  826.5 & 2 & 12 &        \\
835.4 &                 &  833.6 & 0 & 13 & 834.3  \\
856.2 &                 &  859.2 & 1 & 13 &        \\
859.8 &                 &  859.2 & 1 & 13 &        \\
911.4 &                 &  915.1 & 1 & 14 & ?      \\
929.2 & + 11.6 = 940.8  &  938.3 & 2 & 14 &        \\
1009.7& - 11.6 = 998.1  & 1001.3 & 0 & 16 &        \\
1027.1&                 & 1026.9 & 1 & 16 &        \\
1123.3& - 11.6 = 1111.7 & 1113.1 & 0 & 18 & 1114.2 \\
1131.1&                 &        &   &    & Noise? \\
1137.0&                 & 1138.7 & 1 & 18 & 1138.8 \\
1186.0& + 11.6 = 1197.5 & 1194.6 & 1 & 19 &        \\
1192.4&                 & 1194.6 & 1 & 19 &        \\
1234.8& - 11.6 = 1223.2 & 1124.9 & 0 & 20 & 1224.2 \\
1251.8&                 & 1250.5 & 1 & 20 & 1252.0 \\
1265.6& + 11.6 = 1277.2 & 1273.7 & 2 & 20 &        \\
1337.2&                 & 1336.7 & 0 & 22 & 1335.4 \\
1439.0&                 & 1141.4 & 2 & 23 &        \\
1559.5&                 & 1560.3 & 0 & 26 &        \\
\hline
\end{tabular}
\end{center}
\label{FCOMP}
\end{table}

It is interesting to note that 8, out of 11, frequencies determined by us match
well, within a few $\rm\mu Hz$, those listed in the Table 2 of \citet{egge04},
but with a different mode identification.
Therefore we detected signals from the same frequencies as \citet{egge04}, but
extracted the $p$-mode structure in a slightly different way, reaching two
different solutions, and therefore identifications, for the extracted frequencies.

If we use our solution, $\Delta\nu=55.90\,\rm\mu Hz$, 
$\delta\nu_{02}=7.1\,\rm\mu Hz$, $\varepsilon=1.913$, to verify to which 
extent it matches the \citet{egge04} frequencies we find that 16 
frequencies match our solution, 5 need a shift of $+11.6\,\rm\mu Hz$, 
4 need a shift 
of $-11.6\,\rm\mu Hz$, and only 2 do not match the solution. Otherwise, if we
use the \citet{egge04} solution, $\Delta\nu=55.50\,\rm\mu Hz$,
$\delta\nu_{02}=4.9\,\rm\mu Hz$, $\varepsilon=1.537$, we find that 12 
frequencies match this solution, 8 need a shift of $+11.6\,\rm\mu Hz$, 3 
need a shift of $-11.6\,\rm\mu Hz$, and 4 do not match the solution. This 
means that the frequencies detected by \citet{egge04} provide a slightly
better fit to our solution than to their solution. 

This statement is supported by the fact that if we only look at those
12 frequencies that fit, without shift, the \citet{egge04} asymptotic relationship and compare
them with the \citet{egge04} 16 raw  frequencies that fit the solution
found in the present study, we see that the scatter for the 16 frequencies is about 10\% lower 
when we use our asymptotic relation. Moreover,
\citet{egge04} identify the 50\% of those 12 frequencies that fit its relation
as $l=2$ modes, while the
$l=1$ modes should show instead the highest amplitude.

However, since the two solutions provide basically the same frequencies,
one should of course be cautious in claiming that one solution is far
better than the other one.

\subsection{Mode lifetimes \label{LIFETIME}}
Solar-like $p$-mode oscillations in Procyon are presumably excited stochastically by the action of the surface convection \citep {houdek99} as in the case of the Sun's oscillations. Therefore it is interesting to give an estimate of the lifetime of modes for verifying the excitation mechanism that in turn determines their observed spectral width and amplitude. 
A coherent, over-stable, long-lived pulsation mode would concentrate all its power in a small frequency bin, whilst a stochastically, short-lived excited mode would spread its power over a larger frequency interval. 
In the case of the Sun the mode lifetime, at the peak amplitude of $3\,\rm mHz$, is about $3\pm 1\,\rm d$, as estimated by fitting a model to the power distribution for each mode in the PS \citep {chaplin97}.
The \citet {chaplin97} technique requires observations for a significantly longer time than the mode lifetime, that is not necessarily consistent with our observing period of 6 nights. 

\begin{figure}
\centering
\includegraphics[width=8cm]{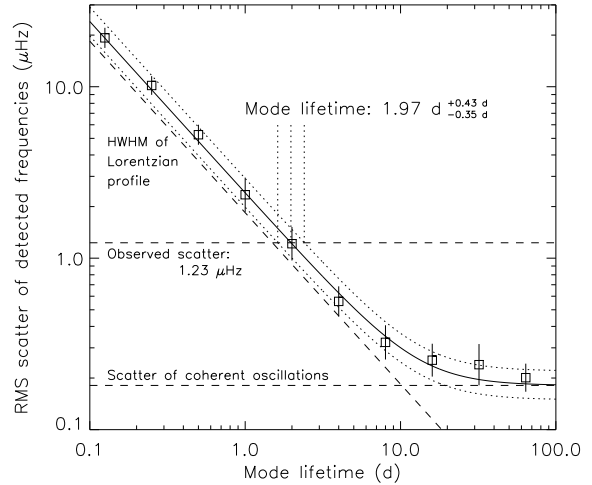}
\caption{The rms scatter of the 10 identified frequencies as a function of mode lifetime obtained by several thousand simulations. The intersection of the horizontal line of the observed scatter with the solid curve obtained by simulations indicates that the mode lifetime is $1.97^{+0.43}_{-0.35}\,\rm d$. The horizontal line of coherent oscillation scatter, corresponding to an infinite mode lifetime, is also indicated. The frequency spread reflects essentially the width of the Lorentzian profile.}
\label{modelife}
\end{figure}

The method we use in the present analysis is based on the frequency scatter per identified mode \citep{bedding04, kjeldsen05}, as calculated in Section \ref{OSCILLATION}. We ran 8400 individual simulations in order to determine the rms scatter of the detected frequencies as a function of the mode lifetime, by using the \citet{DeRidder06} simulator, and neglecting rotational splittings both because Procyon is a slow rotator and the majority of modes used are radial. The simulations contain the mode correct amplitude, noise level, and number of detected modes out of the total number of modes contained in the $p$-mode spectrum, as determined from the asymptotic relationship (\ref{AsymptoticRel}) in the observed frequency interval. The fit represents a model that contains two components: the frequency spread, that is essentially the width of the Lorentzian profile, and the scatter of coherent oscillations, that corresponds to an infinite mode lifetime. The results are reported in Fig. \ref{modelife} where the rms scatter of the $10$ considered frequencies, out of the $11$ identified (see Section \ref{OSCILLATION}), is plotted vs. mode lifetime. Since the mode frequency scatter determined in Section \ref{OSCILLATION} results to be $1.23\,\rm\mu Hz$, it is easily seen from Fig. \ref{modelife} that the corresponding lifetime is $2.0\pm 0.4\,\rm d$. The observed frequency scatter in Procyon is sensibly larger than that consistent with coherent oscillations, ruling out this possibility, and the related mode lifetimes appear to be slightly shorter than those observed in the Sun, where the oscillations are excited by the surface turbulent convection. The present data strongly indicate that oscillations in Procyon are not coherent, but damped and re-excited by a mechanism similar to that operating in the Sun, as expected. 

\begin{figure}
\centering
\includegraphics[width=8cm]{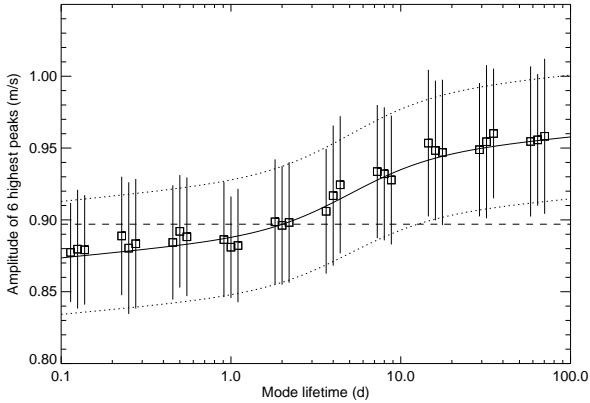}
\caption{The amplitude of the 6 largest amplitude detected modes, for $l=0,1,2$ as a function of the mode lifetime. The bars indicate the $1\,\sigma$ error.}
\label{modelife1}
\end{figure}

The results of the previous method for determining the mode lifetimes are corroborated by a different analysis we performed, based on the amplitude of the highest peaks in the PS, with the idea that also the amplitude of modes are affected by their lifetimes, in the sense that when the peak amplitude is smaller the mode lifetime is shorter, whilst when it is larger the mode lifetime is longer. However, since there is not a simple relationship between amplitude and lifetime, we used a data simulation, based on several thousand runs, that allows a direct comparison between the observed properties of the time-series and the properties of the oscillation modes in order to establish a calibrated relationship between the peak height and the mode lifetime \citep {DeRidder06}. With respect to the previous one, this method has the advantage of being independent of mode identification but the disadvantage of being less accurate.
The result is reported in Fig. \ref{modelife1}, where we show the amplitude, with $1\,\sigma$ error bars, of the 6 largest amplitude detected modes with $l=0,1,2$ as a function of the mode lifetime.  
In Fig. \ref{modelife1} the intersection of the horizontal line corresponding to that of the observed mode of largest amplitude (see Fig. \ref{power}) with the simulated data gives an average mode lifetime $\simeq 2\,\rm d$, in good agreement with the frequency scatter method but with a much larger uncertainty. It is interesting to note the expected trend of mode amplitudes as a function of their lifetimes.

\begin{figure}
\centering
\includegraphics[width=8cm]{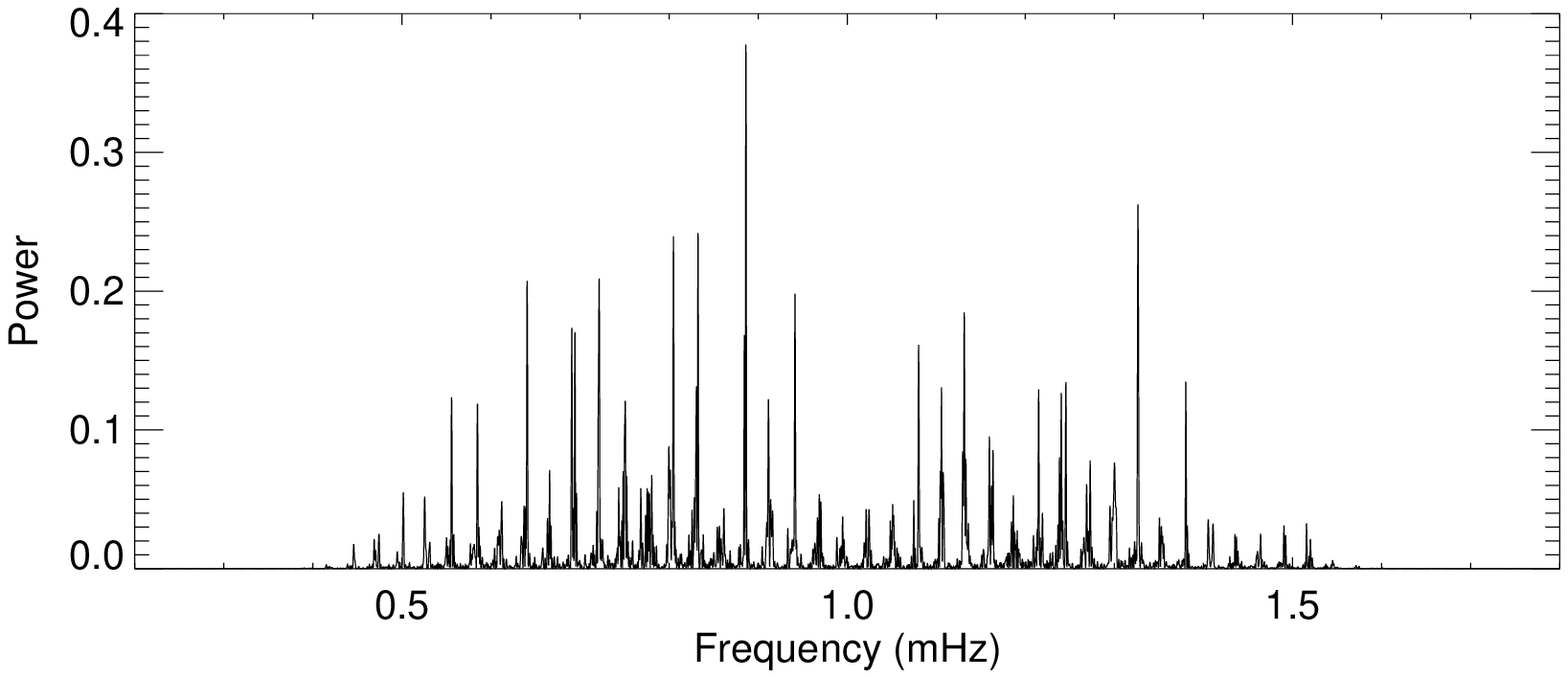}
\includegraphics[width=8cm]{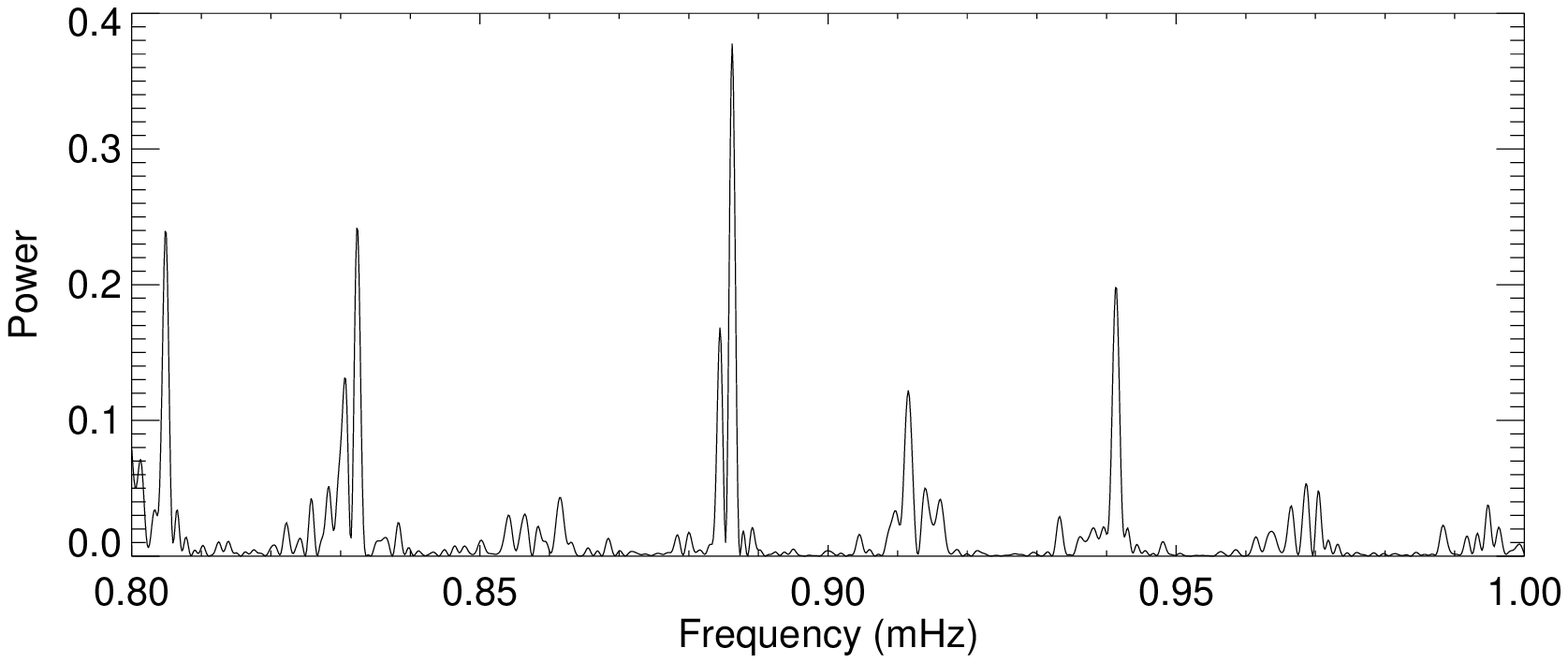}
\caption{Simulated Procyon $p$-mode power spectrum obtained by using the \citet{DeRidder06} simulator not including noise and rotational splittings. The length of the time-series is $10\,\rm d$ and the mode lifetime is $2\,\rm d$.}
\label{sim01}
\end{figure}

\begin{figure}
\centering
\includegraphics[width=8cm]{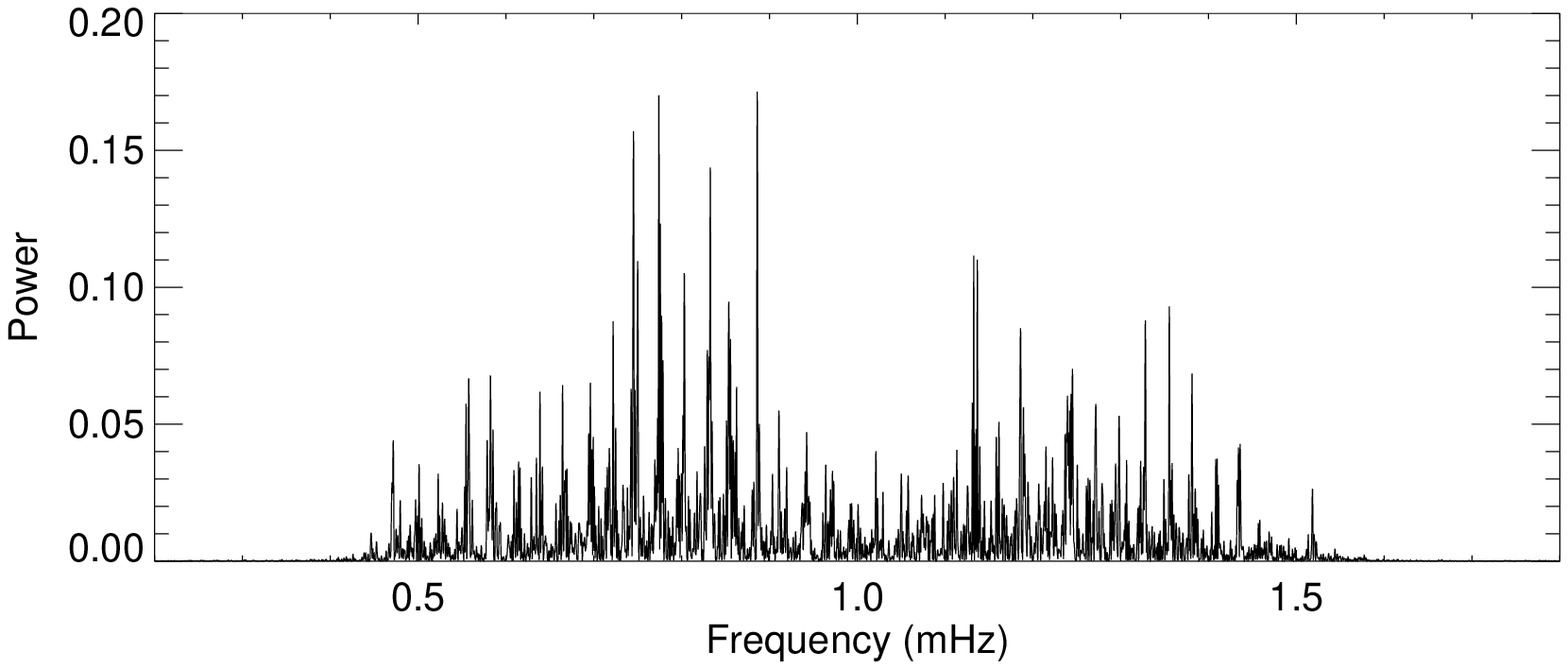}
\includegraphics[width=8cm]{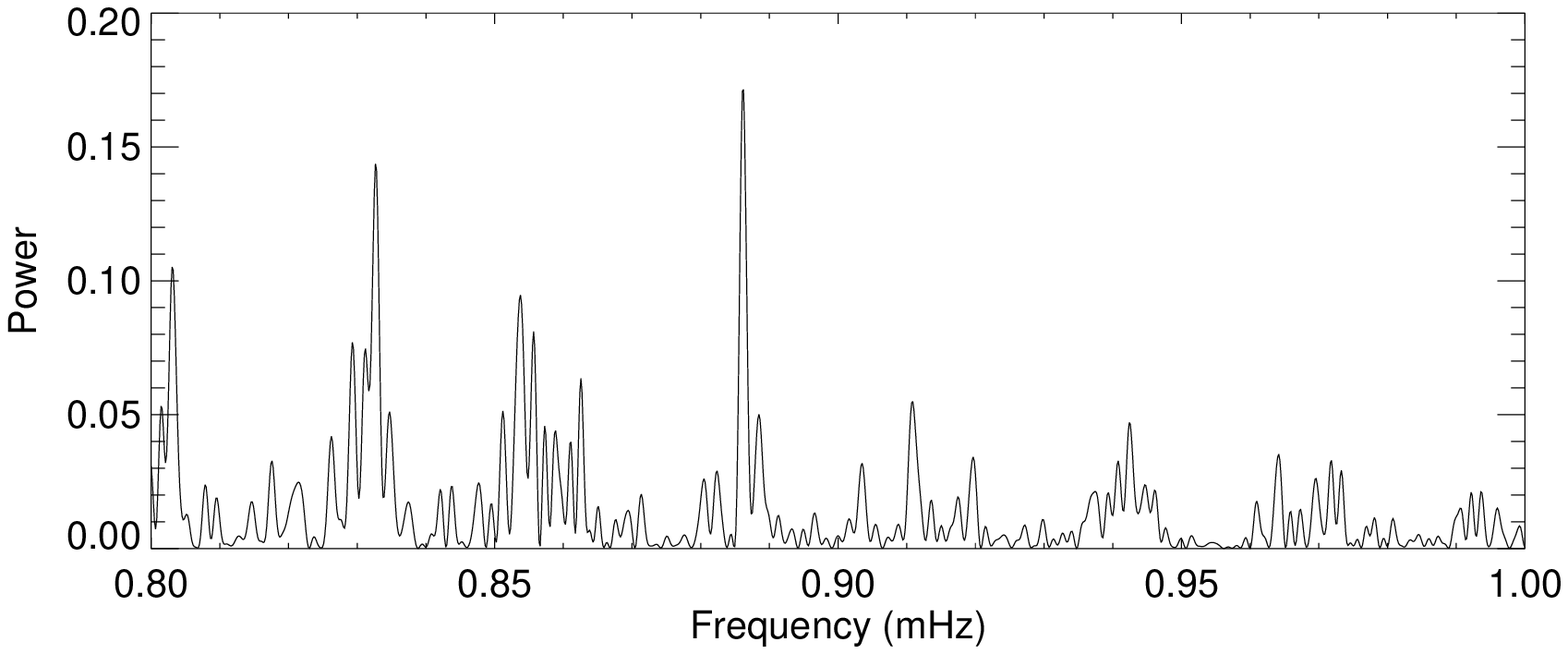}
\caption{The same simulation as that of Fig. \ref{sim01}, but for a mode lifetime of $0.5\,\rm d$. Note that the frequency spreads are larger and peak amplitudes are smaller than those of the modes seen in Fig. \ref{sim01}.}
\label{sim02}
\end{figure}

The effects of frequency scatter and mode amplitude on mode lifetime is evident in Figs. \ref{sim01} and \ref{sim02}, where simulations concerning $10\,\rm d$ continuous observation campaigns are reported for mode lifetimes of $2\,\rm d$ and $0.5\,\rm d$ respectively. In the above quoted two figures the upper panel shows the complete PS, while the lower one an enlarged portion of it from $0.8\,\rm mHz$ to $1.0\,\rm mHz$. From our simulations it is clearly seen that the frequencies of long-living modes are much better defined, or less scattered, and have larger amplitudes than those of short-living modes.

\section{Line equivalent width measurements and analysis \label{EQWIDTH}}
As already mentioned, the red part of the echelle spectrum of Procyon is insensitive to the iodine cell calibration
and therefore the search for stellar pulsations can be performed only by means of measurements of line equivalent widths (EW). The presence of telluric lines and limited number of strong lines sensitive to temperature
(e.g. $\rm H_\alpha$ is located at the border of spectral orders and is
therefore useless for accurate EW measurements)
restricted our analysis to the spectral lines reported in Table \ref{SpectralLines}.
\begin{table}[h]
\caption[]{Spectral line wavelengths used for EW measurements.}
\begin{center}
\begin{tabular}{cc}
\hline
Wavelength (\AA) & Spectral lines\\
\hline
6393.4& Fe I\\
6436.7& Ca I\\
6462.4& Ca I / Fe I\\
6546.0& Fe I\\
6574.9& Fe I\\
6592.7& Fe I\\
6643.4& Ni I\\
6677.8& Fe I\\
6717.5& Ca I\\
\hline
\end{tabular}
\end{center}
\label{SpectralLines}
\end{table} 

\subsection{Power spectrum \label{EWPOWER}}
\begin{figure}
\centering
\includegraphics[width=8cm]{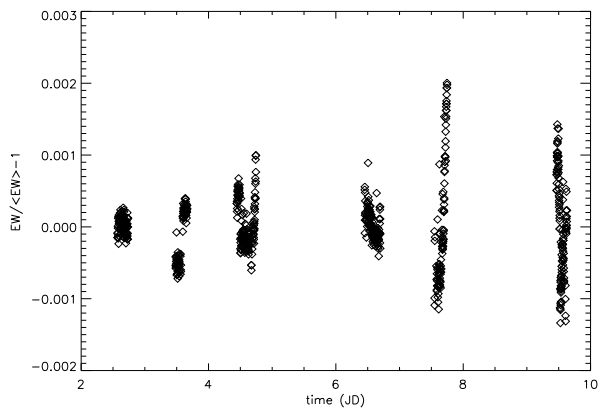}
\caption{Time series of EW measurements as derived from the combination of all
the lines reported in Table \ref{SpectralLines}.}
\label{EWtimeSeries}
\end{figure}

\begin{figure}
\centering
\includegraphics[width=8cm]{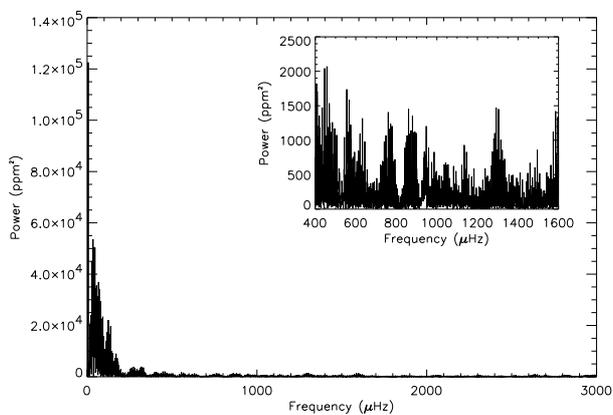}
\caption{The weighted power spectrum $\rm EW_{\rm best}-<\rm EW_{\rm
best}>$ derived from the combination of the measurements of all the lines of
Table \ref {SpectralLines}. There is no excess of power in the range $0.5 - 1.5\,\rm mHz$, but the
background power in the range $0.2 - 0.7\,\rm mHz$ could be a signature of stellar granulation.}
\label{Power2}
\end{figure}

Since the expected variations in the EW induced by the stellar pulsations are of the order of a few ppm, we
adopted an accurate approach for the measurement of the EW, following
\citet{kjeldsen95}. Instead of fitting the line profile for determining the EW, we computed 
the flux in three artificial filters, analogously to Str\"omgren $\rm H_\beta$ photometry. The first filter was centred on the considered line at the wavelength $\lambda_0$, with a band-width $\Delta\lambda_0$, and the other two on the near continuum with respect to the line at $\lambda_0$, one placed on the red
region at $\lambda_{\rm r}$, with band-width $\Delta\lambda_{\rm r}$, and the other on the blue one at $\lambda_{\rm b}$, with a band-width $\Delta\lambda_{\rm b}$.
We then computed the EW, or more precisely the line-index, in terms of the radiation flux measured in the selected filters ${\cal F}(\Delta\lambda)$: 
\begin{equation} 
\rm EW=\left[\frac{{\cal F}(\Delta\lambda_0)}{\Delta\lambda_0}\right]\cdot\left[\frac{\Delta\lambda_{\rm r}+\Delta\lambda_{\rm b}}{{\cal F}(\Delta\lambda_{\rm r})+{\cal F}(\Delta\lambda_{\rm b})}\right]                                                          
\label{EWformula}
\end{equation}
for two different central filter widths, $\Delta\lambda^1_0$ and $\Delta\lambda^2_0$, in order
to obtain two values for the EW, $\rm EW_1$ and $\rm EW_2$, from which the quantity ${\rm EW_{\rm best}}= {\rm EW}_1^{m+1}/ {\rm EW}_2^m$ can be constructed. 
The aim of this procedure is to construct a quantity that is less sensitive to
the continuum fluctuations, determining $m$ so as to minimize the
scatter of the measurements in a similar way as 
described for the radial velocity analysis. The procedure was repeated for all
the lines reported in Table \ref{SpectralLines}, and then, in order to improve the
S/N, we combined all the $\rm EW_{\rm best}$'s obtained for the examined lines,
assuming for them the same sensitivity to temperature changes, and  derived a weighted mean value.
In Figs. \ref{EWtimeSeries} and \ref{Power2} we show the time-series and
the weighted PS of the $\rm EW_{\rm best}$, respectively.
The weights adopted are $w_i=1/{\sigma_i^2}$, where the $\sigma_i$'s are the standard deviations of the data. 

\subsection{Granulation noise \label{GRANULATION}}
The combined power spectra do not show evidence of power excess in the range $0.5 - 1.5\,\rm mHz$, as is clear
from Fig. \ref{Power2}. However, at low frequencies, some concentration of
power, similar to the background power reported by \citet {kjeldsen99}
for $\alpha\,\rm Cen\, A$, and attributed to stellar granulation, is apparent.
Analogously to \citet {kjeldsen99}, here we attempt to give an estimate
of the amount of power related to Procyon granulation.
In this procedure we optimize the S/N in the PS by computing the
weighted means of the measurements, by using weights that minimize the mean noise for each line in the interval
$0.2 - 0.7\,\rm mHz$, and then rescale the weights for each line in such a way that the weights, $w_i$, satisfy the relationship $\sum w_i=1/\cal A$, where  
$\cal A$ is the mean of the peak amplitudes in the range $0.2 - 0.7\,\rm mHz$.

\begin{figure}
\centering
\includegraphics[width=8cm]{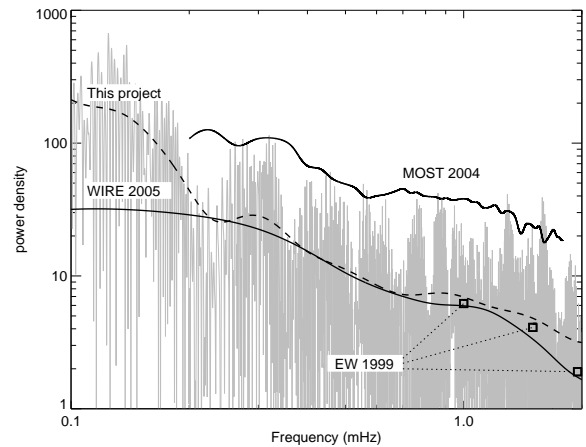}
\caption{Smoothed power density spectra (PDS) of Procyon granulation noise as obtained from SARG (dashed line, this project) and AAT/ESO [EW 1999, squares, \citep {kjeldsen99}] spectroscopy, and MOST \citep{matthews04} and WIRE \citep{bruntt05} space photometry (continuous lines). On background the PDS derived from the PS of Fig. \ref{Power2} is shown. Power density is expressed in $\rm ppm^2\,\mu Hz^{-1}$.}
\label{PowerDensity}
\end{figure}

In order to evaluate the power generated by the granulation of the star in the
PS of the EW's and make a comparison with data obtained with different
methods we need to define a quantity independent of the temporal length of the data, that is obtained by computing the power density spectrum (PDS), a measure of the power per frequency resolution element that
is therefore independent of the length and sampling of the time-series:
\begin{equation}
\rm PDS(\nu)={\rm A^2(\nu)}\left[{\int_0^\nu \rm N(\nu)W^2(\nu)d\nu}\right]^{-1}
\end{equation}
where $\rm A(\nu)$ is the amplitude (in ppm) in the PS, $\rm W(\nu)$
the window function, and $\rm N(\nu)$ the Nyquist frequency.
The result of our analysis is shown in Fig. \ref{PowerDensity}, where the dashed line (this project) indicates the smoothed power density spectrum of Procyon granulation noise, and background the PDS of the PS shown in Fig. \ref{Power2}.

\subsection{Comparison with other granulation noise determinations \label{COMPARISON_NOISE}}
The literature reports three previous claims of granulation noise power detection in Procyon. The first is by \citet {kjeldsen99},
who found a granulation noise slightly above the solar level
in the frequency interval $1-2\,\rm mHz$, by using the Balmer EW's. These determinations are shown in Fig. \ref{PowerDensity} as squares (EW 1999). The other two claims, instead of spectroscopic, concern space photometric measurements carried out by the satellites MOST \citep{matthews04}, operating in the bandwidth $3700-7000\,\AA$, and WIRE \citep{bruntt05}, operating with the white light star tracker. 

While our measurements are directly comparable with the spectroscopic ones of \citet{kjeldsen99}, in order to compare them with the photometric ones, we need to pass from intensities, $I$, to temperatures, $T$, and from the latter to EW's. This is accomplished by using appropriate sensitivity indices; for intensity we used  $\partial\ln I/\partial\ln T = 4.5$ \citep{KjeldsenBedding95}, and for EW, $\partial\ln\rm (EW)/\partial\ln T = -7.8$, as deduced from the analysis of \citet{bedding} for Fe I lines at $T = 6500\,\rm K$. The results are shown in Fig. \ref{PowerDensity}, where MOST and WIRE PDF curves are plotted together with \citet{kjeldsen99} data and our curve. 

It appears that, while WIRE and our distribution of power are mutually consistent, at least in the frequency interval $0.2 - 2\,\rm mHz$, and corroborated by the few data of \citet{kjeldsen99}, MOST shows a power density level higher by a factor of four or more. Uncertainties in sensitivity index calibration of $I$ and EW with respect to $T$ may lead to some discrepancies, but not so large to explain such a significant disagreement between MOST and other data. In order to render the MOST data comparable with ours, it should be $\mid\partial\ln\rm (EW)/\partial\ln T\mid\; < 4.5$ and $\partial\ln I/\partial\ln T > 4.5$. This is unlikely because the value of $\partial\ln\rm (EW))/\partial\ln T$ adopted here has been derived by observations of the same lines we use here at the same temperature of Procyon \citep{bedding} and a value of 
$\partial\ln I/\partial\ln T > 4.5$ is in disagreement with the models of \citet{bedding}. This seems to imply that MOST did not detect granulation noise, probably because of some instrumental or stellar spurious signal as suggested by \citet{bruntt05}.  

Differently from the first spectroscopic determinations of \citet{kjeldsen99} relative to a few points in the high frequency side of the granulation noise PDS, our EW determinations constitute the first spectroscopic measurements of Procyon granulation in the frequency range $0.1 - 2.0\,\rm mHz$, so providing an independent measure of and establishing an upper limit to granulation noise power. The fact that at low frequencies, below $0.2\,\rm mHz$, our power level is higher than that shown by WIRE might be caused partially because we did not correct the WIRE data at low frequencies for the effect of high-pass filter and partially  because of the possible effect of stellar activity in Procyon that affects spectral lines but not the continuum.   

\section{Conclusions \label{CONCLUSIONS}}
The analysis of the PS of Procyon obtained by SARG high resolution spectrograph shows unequivocally the presence of a solar-like $p$-mode spectrum as clearly demonstrated in Fig. \ref{foldedspectrum}. The main seismic properties of Procyon are summarized in Table \ref{PROPERTIES}.
\begin{table}[h]
\caption[]{Properties of the $p$-mode oscillations in Procyon.}
\begin{center}
\begin{tabular}{lc}
\hline
Large separation $\Delta\nu$ & $55.90\pm 0.08\,\rm\mu Hz$ \\
Small separation $\delta\nu_{02}$ & $7.1\pm 1.3\,\rm\mu Hz$ \\
Asymptotic relationship $\varepsilon$ & $1.913\pm 0.025$ \\
Frequency scatter per mode & $1.23\,\rm\mu Hz$ \\
$\nu(l=0,n=16)$ & $1001.3\pm 0.4\,\rm \mu Hz$ \\
Mean amplitude per mode ($l=0,1$) & $0.45\pm 0.07\,\rm m\,s^{-1}$ \\
Mode lifetime &  $2.0\pm 0.4\,\rm d$ \\
\hline
\end{tabular}
\end{center}
\label{PROPERTIES}
\end{table}

We identified with a good margin of certainty $11$ individual $p$-mode frequencies with $l=0,1,2$ and $7\leq n \leq 22$ in the frequency range $500-1400\,\rm\mu Hz$, as reported in Table \ref{FI},
which are largely consistent with those found by \citet{egge04}. 

The large frequency separation deduced from the fit of $10$, out of the $11$ identified frequencies, to the asymptotic relationship (\ref{AsymptoticRel}) is consistent with the results from the comb response analysis, both individual and cumulative, and its value agrees fairly well with that determined by \citet{egge04}, $\Delta\nu = 55.5\,\rm\mu Hz$, but shows some discrepancy with that determined by \citet{martic04}, $\Delta\nu = 53.6\,\rm\mu Hz$,
though in a previous article \citep{martic99} the authors found $\Delta\nu = 55.6\,\rm\mu Hz$, as is evident from their Fig. 12. 

The large frequency separation we have determined is consistent with an evolutionary model of Procyon, constructed with solar chemical abundance, elemental diffusion and mass loss, having a mass of $1.48\,\rm M_\odot$, an age of $1.8\,\rm Gy$, a luminosity of $6.9\,\rm L_\odot$, a radius of $2.04\,\rm R_\odot$, and a surface convection zone deep the $8\,\%$ of the stellar radius \citep{bonanno06}.
It is in very good agreement with other
theoretical studies of Procyon A \citep{barban99, chaboyer99, egge05}.
  
The small frequency separation we determined from one $l=2$ and
three $l=1$ modes is larger by $1.6\,\sigma$ than that determined by \citet{martic04}, $\delta\nu_{02}=5.1\,\rm\mu Hz$, and that we derived from \citet{egge04} data, $\delta\nu_{02}=4.9\,\rm\mu Hz$, but our value is not strongly constrained by our measurements, as only a few modes with $l\neq 0$ were identified.

The mean amplitude per mode, for modes with $l=0,1$, is about $0.45$, twice larger than the solar one. 

The mode lifetime has been determined by means of two independent methods whose results are mutually consistent and give a lifetime of about $2\,\rm d$.    
This short lifetime rules out the possibility of a coherent, over-stable excitation of modes in Procyon, but indicates that oscillations are excited stochastically by turbulent convection as in the case of the Sun.   

We used the red part of the observed spectrum for determining, through the measurement of the EW's, an upper limit to granulation noise power that resulted to be in agreement with previous spectroscopic \citep{kjeldsen99} and space photometric  \citep{bruntt05} measurements, but in significant contrast with MOST data \citep{matthews04}.

Our radial velocity data are available upon request to the first author of the present article.

\begin{acknowledgements}
This work has partially been supported by the Italian Ministry of Education, University and
Scientific Research under the contract PRIN 2004024993.
\end{acknowledgements}

\bibliographystyle{aa}

\end{document}